# Habitat variability does not generally promote metabolic network modularity in flies and mammals

Kazuhiro Takemoto[*]

*Department of Bioscience and Bioinformatics, Kyushu Institute of Technology, Iizuka, Fukuoka 820-8502, Japan*
* Corresponding author: E-mail: takemoto@bio.kyutech.ac.jp

## Abstract

The evolution of species habitat range is an important topic over a wide range of research fields. In higher organisms, habitat range evolution is generally associated with genetic events such as gene duplication. However, the specific factors that determine habitat variability remain unclear at higher levels of biological organization (e.g., biochemical networks). One widely accepted hypothesis developed from both theoretical and empirical analyses is that habitat variability promotes network modularity; however, this relationship has not yet been directly tested in higher organisms. Therefore, I investigated the relationship between habitat variability and metabolic network modularity using compound and enzymatic networks in flies and mammals. Contrary to expectation, there was no clear positive correlation between habitat variability and network modularity. As an exception, the network modularity increased with habitat variability in the enzymatic networks of flies. However, the observed association was likely an artifact, and the frequency of gene duplication appears to be the main factor contributing to network modularity. These findings raise the question of whether or not there is a general mechanism for habitat range expansion at a higher level (i.e., above the gene scale). This study suggests that the currently widely accepted hypothesis for habitat variability should be reconsidered.

## 1. Introduction

The evolution of species habitat range is an important topic of scientific inquiry at several scales of biological research, from fundamental biological processes to ecology (Bridle & Vines, 2007; Root et al., 2003; Roy, Hunt, Jablonski, Krug, & Valentine, 2009), particularly in the context of predictions related to biodiversity and climate change. Therefore, understanding the factors that determine species habitat use is a relevant topic for advancing these research fields. In particular, it is important to identify the molecular (microscopic) mechanisms that contribute to determining a species habitat range, because the behavior of a species (macroscopic) may result from complex biological systems. For example, some previous studies (Barrett & Schluter, 2008; Kellermann, van Heerwaarden, Sgrò, & Hoffmann, 2009) suggested the importance of genetic variation to the ability to adapt and exploit new environments. Moreover, recent studies reported that gene duplication promotes habitat variability in flies (*Drosophila* species) (Makino & Kawata, 2012) and mammals (Tamate, Kawata, & Makino, 2014). This work was inspired by the proposed importance of gene duplication to increasing biological robustness and evolvability (Wagner, 2008), which are themselves related to habitat variability.

However, it remains unclear how these genetic events affect species habitat variability at a higher level of biological organization. In this context, evaluation of modular organization in biological systems (Hartwell, Hopfield, Leibler, & Murray, 1999) is useful because it is also generally considered to be related to robustness (Hintze & Adami, 2008) and evolvability (Yang, 2001), despite some opinions to the contrary(Hansen, 2003; Holme, 2011). The evolution of modularity in cellular networks has been specifically intriguing to researchers in the context of network biology(Barabási & Oltvai, 2004; Takemoto, 2012a). In particular, a hypothesis has been proposed that variability in natural habitats promotes network modularity. For example, in a theoretical model, Kashtan and Alon (Kashtan & Alon, 2005) showed that modular networks spontaneously evolved when a fitness peak determined by the environment changes over time in a manner that preserves the same subgoals but in different permutations. Similarly, Lipson et al. (Lipson, Pollack, & Suh, 2002) suggested that changing environments could promote modularity. Hintze and Adami (Hintze & Adami, 2008) showed that modularity evolves in biological networks (modeled as metabolic networks) to deal with a multitude of functional goals, with the degree of modularity depending on the extent of environmental variability.

In this context, metabolic networks are particularly interesting because metabolic processes are essential for physiological functions and for maintaining homeostasis in living organisms (Takemoto & Oosawa, 2012; Takemoto, 2012a). Metabolic networks also determine the behavior of organisms, such as the space use (Jetz, Carbone, Fulford, & Brown, 2004) and feeding rate (Brown, Gillooly, Allen, Savage, & West, 2004) of animals, which may in turn be related to habitat variability. In addition, analyses can be performed using actual empirical data, because metabolic networks are available for a wide diversity of species in databases such as the Kyoto Encyclopedia of Genes and Genomes (KEGG) (Kanehisa et al., 2014) and the Encyclopedia of Metabolic Pathways (MetaCyc) (Caspi et al., 2012). In fact, using network analysis, Parter et al. (Parter, Kashtan, & Alon, 2007) showed that variability in natural habitats promotes the modularity observed in metabolic networks. This result clearly supports the predictions derived from theoretical models, and several researchers have actively investigated the ecological interactions underlying metabolic networks according to habitat variability (Chave, 2013; Levy & Borenstein, 2012).

Despite this recent attention to this relationship between modularity and habitat viability, more comprehensive examinations are required to resolve some outstanding questions. Indeed, recent studies have cast doubt on this relationship. For example, several alternative theories for explaining the origin and evolution of modularity have been proposed, including the neutral theories of protein (Solé & Valverde, 2008) and metabolic networks (Takemoto, 2012b), connection-cost theory (Clune, Mouret, & Lipson, 2013), and multiplicative-mutation theory (Friedlander, Mayo, Tlusty, & Alon, 2013). Furthermore, a study conducted in archaea, a type of prokaryote distinct from bacteria, did not find a positive correlation between habitat variability and metabolic network modularity (Takemoto & Borjigin, 2011). Similarly, in bacteria, no positive correlation was observed using the latest version of the metabolic database (Takemoto, 2013; Zhou & Nakhleh, 2012). In short, the observed associations between metabolic network modularity and habitat variability (Parter et al., 2007) may be the result of an artifact due to lack of available data on metabolic reactions. More importantly, the studies conducted thus far are limited to lower organisms such as bacteria and archaea.

Therefore, the aim of this study was to investigate the relationship between habitat variability and metabolic network modularity in higher organisms, including data from flies (Makino & Kawata, 2012) and mammals (Tamate et al., 2014). In addition to the potential effect on habitat variability in promoting network modularity, an association between gene duplication and habitat variability has also been observed. Given that gene duplication also influences the metabolic network structure (Barabási & Oltvai, 2004; Díaz-Mejía, Pérez-Rueda, & Segovia, 2007; Papp, Pál, & Hurst, 2004; Takemoto, 2012a), it is reasonable to hypothesize that habitat variability may be linked to not only gene duplication but also metabolic network modularity. To investigate these relationships, data related to habitat variability were collected from the published literature; data were collected only from species for which metabolic network data are also available (see Section 2). Using these data, I evaluated whether habitat variability increases metabolic network modularity and how the association between gene duplication and habitat variability might influence the modularity of the metabolic network.

## 2. Material and Methods

### *2.1 Collection of data related to habitat variability and fraction of duplicated genes*

The data on habitat variability and fraction of duplicated genes in flies and mammals were obtained from Makino and Kawata (Makino & Kawata, 2012) and Tamate et al.(Tamate et al., 2014), respectively. Habitat variability was measured based on the Köppen climate classification in habitat areas for living organisms (see (Makino & Kawata, 2012; Tamate et al., 2014) for details). In this study, the Brillouin index was used for measuring habitat variability, because such indices tend to follow a normal distribution for species, which is important for statistical analyses. Previous studies have considered two definitions of habitat variability: the Brillouin index and climate envelope. However, similar conclusions were generally obtained using these two definitions, because of a strong positive correlation between them (Supplementary Tables S1 and S2).

For this analysis, species were selected based on the availability of metabolic network data in the KEGG database (Kanehisa et al., 2014). Finally, data were collected for 11 fly species (*Drosophila* spp.) (Supplementary Table S1) and 14 mammalian species (Supplementary Table S2).

## 2.2 Construction of metabolic networks

The procedure for metabolic network construction is generally the same as that reported previously (Takemoto, 2013). XML files (version 0.7.1) containing the metabolic network data were downloaded from the KEGG database (Kanehisa et al., 2014) (ftp://ftp.genome.jp/pub/kegg/xml/kgml/metabolic/organisms/) on August 11, 2015, and two types of metabolic networks were constructed: compound networks and enzymatic networks. As of July 1, 2011, the KEGG FTP site is only available to paid subscribers. Since the use of such data may be desirable to ensure reproducibility, the present dataset on metabolic networks is available upon request.

Compound networks are represented as directed networks, in which the nodes and edges correspond to metabolites and reactions (i.e., substrate–product relationships), respectively. R numbers (e.g., R00010) were extracted from the XML files, which indicate metabolic reactions. On the basis of the R numbers, substrate–product relationships and reversibility/irreversibility of chemical reactions were identified as carbon traces using the metabolic reaction database (Stelzer, Sun, Kamphans, Fekete, & Zeng, 2011). Currency (ubiquitous) metabolites such as $H_2O$, ATP, and NADH were removed, as described previously (Takemoto, Nacher, & Akutsu, 2007).

Enzymatic networks are also represented as directed networks, in which the nodes and edges are metabolic enzymes (reactions) and the presence of interjacent chemical compounds, respectively. In brief, an edge is drawn between 2 enzymes (nodes) if at least 1 product of a reaction catalyzed by an enzyme corresponds to at least 1 substrate of the reaction catalyzed by another enzyme (see (Takemoto, Niwa, & Taguchi, 2011; Takemoto, 2012a) for details). Substrate–product relationships and reversibility/irreversibility of chemical reactions were defined as carbon traces using the metabolic reaction database (Stelzer et al., 2011) in order to avoid the emergence of biologically unsuitable edges (see (Takemoto et al., 2011; Takemoto, 2012a) for details of the importance of this handling procedure and an example).

The largest (weakly) connected component (giant component) was extracted from each metabolic network to accurately calculate network modularity, and for comparison with the previous study (Parter et al., 2007). In particular, network modularity may be overestimated or underestimated due to isolated components (Parter et al., 2007; Takemoto & Borjigin, 2011; Takemoto, 2013); thus, the use of entire networks was avoided in the present study. However, similar conclusions were also obtained using entire networks.

## 2.3 Network modularity

The modularity of networks is often measured using the $Q$-value (reviewed by Fortunato (Fortunato, 2010)). $Q$ is defined as the fraction of edges that lie within, rather than between, modules relative to that expected by chance (see Equations (14) and (37) in Fortunato (Fortunato, 2010) for the definitions of $Q$s for undirected networks [$Q_d$] and directed networks [$Q_{ud}$], respectively).

A network with a higher $Q$-value indicates a higher modular structure. Thus, the global maximum $Q$ over all possible divisions should be identified. Since it is hard to find the optimal division with the maximum $Q$ in general, approximate optimization techniques are required (Fortunato, 2010). In this study, spectral decomposition methods were used for undirected networks (Newman, 2006) and directed networks (Leicht & Newman, 2008) in order to avoid the resolution limit problem in community (or module) detection (Fortunato & Barthélemy, 2007; Fortunato, 2010) as much as possible (and to facilitate comparison with previous studies (Clune et al., 2013; Parter et al., 2007)). The maximum $Q$-value is thus defined as the network modularity.

In general, the normalized network modularity value ($Q_m$), calculated by comparing the $Q$-value of an empirical network with the mean $Q$-value of randomized (null model) networks, is used, which allows for comparison of network modularity among networks of variable size and connectivity (Parter et al., 2007; Takemoto & Borjigin, 2011; Takemoto, 2013). However, $Q_m$ was not considered in this study because network size and connectivity are similar between species of flies (Supplementary Table S1) and mammals (Supplementary Table S2). In addition, the use of randomized networks can cause problems. For example, Artzy-Randrup et al. (Artzy-Randrup, 2004) and Beber et al.(Beber et al., 2012) reported that network measures were overestimated due to the use of biologically implausible null model (randomized) networks. However, similar conclusions were also obtained using $Q_m$ values calculated based on randomized networks that were generated using a simple edge-switching algorithm (Milo et al., 2002). Therefore, this limitation does not affect the conclusions.

## 2.4 PICs

To remove any phylogenetic effects on the association between biological variables, PICs of the variables were computed from phylogenetic trees, using the function *pic* in the R-package *ape* (version 3.3). A robust phylogenetic tree of flies (*Drosophila* spp.) was obtained from the *Drosophila* 12 Genomes Consortium (Clark et al., 2007), according to a previous study (Makino & Kawata, 2012) (Supplementary Data S1). The phylogenetic tree of mammals was constructed using the matrix extracellular phosphoglycoprotein precursor (*MEPE*) gene (Bardet, Delgado, & Sire, 2010), according to a previous study(Tamate et al., 2014). In particular, the *MEPE* gene sequences of 13 mammal species were obtained from the KEGG database (Kanehisa et al., 2014) on August 19, 2015. Since the *MEPE* gene is not found in the platypus *Ornithorhynchus anatinus* (Bardet et al., 2010), this species was omitted from this analysis. After the multiple alignments of the 13 *MEPE* nucleotide sequences using ClustalW2 software (Larkin et al., 2007), the phylogenetic tree was constructed (Supplementary Data S2) using NJplot (doua.prabi.fr/software/njplot).

## 2.5 Statistical analyses

To measure the statistical dependence between biological variables (considering PICs), Pearson's correlation coefficient ($r$) and its associated *p*-value were computed using R software (version 3.2.2) (www.r-project.org). Similar conclusions were obtained using

Spearman's rank correlation coefficient (a non-parametric measure, which is relatively robust to outliers and can be used to analyze nonlinear relationships).

ANCOVAs were performed to evaluate the influences of biological variables on metabolic network modularity (considering PICs). In particular, the function *lm* in R software was used.

## 3. Results

### 3.1 Habitat variability promotes compound network modularity in flies and mammals

After the data collection and integration procedures, data on habitat variability, the fraction of duplicated genes, and metabolic networks were obtained for 11 different fly and 14 mammal species (see Section 2.1 and Supplementary Tables S1 and S2).

For these living organisms, compound metabolic networks were constructed, which are represented as directed networks in which the nodes and edges correspond to metabolites and reactions (i.e., substrate–product relationships), respectively (see Section 2.2).

The network modularity of these metabolic networks was computed using the $Q$-value, which is generally defined as the fraction of edges that lie within, rather than between, modules relative to that expected by chance (see Section 2.3). Both the directed version ($Q_d$) and undirected version ($Q_{ud}$) of the $Q$-value was considered, to allow for effective comparison with previous studies. $Q_d$ is suitable for analyzing metabolic networks because the networks are directed (Clune et al., 2013). For example, unlike $Q_{ud}$, $Q_d$ can characterize input (i.e., nutrient) modules, core modules, and output (i.e., product) modules (Leicht & Newman, 2008) (e.g., a bow-tie structure, which is related to biological robustness (Friedlander, Mayo, Tlusty, & Alon, 2014; Ma & Zeng, 2003)) in metabolic networks by considering edge direction. Nonetheless, some previous studies (Parter et al., 2007; Takemoto & Borjigin, 2011; Takemoto, 2013; Wagner & Fell, 2001) focused on the undirected version of the $Q$-value because of some apparent problems with $Q_d$; for example, the reversibility/irreversibility of metabolic reactions may change with environmental conditions (Parter et al., 2007; Wagner & Fell, 2001), and $Q_d$ cannot distinguish between situations with and without directed flow (Arenas, Duch, Fernández, & Gómez, 2007) (e.g., Fig. 14 in Fortunato (Fortunato, 2010)). Since both $Q_d$ and $Q_{ud}$ have different advantages and disadvantages, both values were used to evaluate the robustness of the results.

Although the $Q$-value is generally affected by the network size (i.e., the number of nodes) and the number of edges (Guimerà, Sales-Pardo, & Amaral, 2004), in this study, these network parameters did not influence $Q_d$ and $Q_{ud}$ because they are similar between species of flies and mammals (see Methods and Supplementary Tables S1 and S2). For example, in mammals, the $Q_d$ value of compound networks was not correlated with either network size (Pearson's product correlation coefficient $r = 0.0099$, $p = 0.97$) or the number of edges ($r = -0.0018$, $p = 1.0$), and was also not correlated with genome size ($r = 0.0073$, $p = 0.98$) and the number of protein-coding genes ($r = 0.034$, $p = 0.91$). Similar conclusions were obtained with respect to the $Q_{ud}$ value of compound networks, and for the $Q_d$ and $Q_{ud}$ values of enzymatic networks (Supplementary Tables S1 and S2). These same results were obtained when considering Spearman's rank correlation coefficient (see Section 2.5).

In contrast to expectation based on the hypothesis that habitat variability increases network modularity, no positive correlation was observed between $Q_d$ and habitat variability in the compound networks (Figure 1 and Table 1). To perform more comprehensive examinations, the effect of the phylogenetic relationship was considered. The importance of a phylogeny for evaluating associations among biological parameters is well known in terms of comparative phylogenetic analysis (Felsenstein, 1985; T. Garland, Harvey, & Ives, 1992; Theodore Garland, Bennett, & Rezende, 2005). Such analyses generally assume a simple evolutionary model in which random Brownian motion-like traits are considered to change on a phylogenetic tree with accurate branch lengths. In particular, the phylogenetically independent contrasts (PICs) were computed for these parameters (see Section 2.4). However, the results were not changed when the PICs were considered, in that there was still no positive correlation between network modularity habitat variability. Similar conclusions were also obtained when using $Q_{ud}$ (Figure 2 and Table 1).

## *3.2 No general association between network modularity and habitat variability in enzymatic networks, despite positive associations in flies*

Enzymatic networks were also considered, which are represented as directed networks in which the nodes and edges are metabolic enzymes (reactions) and the presence of interjacent chemical compounds, respectively (see Section 2.2). As for the compound networks described above, the relationship between network modularity and habitat variability in the enzymatic networks was investigated (Table 2). Overall, there was no association between enzymatic network modularity and habitat variability (considering both $Q_d$ and $Q_{ud}$); however, some exceptions were found.

$Q_d$ showed a weakly positive correlation with habitat variability in mammals. However, this positive relationship was only observed when the PICs were not considered. Thus, the observed positive correlation between $Q_d$ and habitat variability might have been an artifact caused by the phylogenetic relationship among mammals.

By contrast, $Q_d$ was negatively correlated with habitat variability in flies when the PICs were considered. However, this correlation might also be an artifact caused by some outliers, because no correlation was found when using Spearman's rank correlation coefficient ($r_s = -0.25$, $p = 0.49$).

In support of the hypothesis that habitat variability promotes network modularity, a positive relationship between $Q_{ud}$ and habitat variability was observed in flies when considering the PICs (Table 2 and Figure 3). In this case, the outliers did not affect the observed correlation, because a positive correlation was also found when using Spearman's rank correlation coefficient ($r_s = 0.78$, $p = 0.012$ when PICs were considered).

## 3.3 Enzymatic network modularity is related to the fraction of duplicated genes rather than to habitat variability in flies

The results described above indicated that more careful examinations are required to resolve the relationship between network modularity and habitat variability. In particular, it was necessary to evaluate whether the observed positive association between $Q_{ud}$ and habitat variability in the enzymatic networks of flies (Figure 3) is a unique finding (i.e., whether the observed correlation is independent of gene duplication, which is also associated with habitat variability).

Thus, the correlation between $Q_{ud}$ and the fraction of duplicated genes was investigated, and a strong positive correlation was found (Figure 4). Figures 3 and 4 show that both habitat variability and the fraction of duplicated genes are related to $Q_{ud}$.

Analyses of covariance (ANCOVAs; see Section 2.5) were performed to evaluate the independent influences of habitat variability and the fraction of duplicated genes on $Q_{ud}$, in order to determine which of the factors is the main determinant of $Q_{ud}$. The results showed that $Q_{ud}$ is mostly determined by the fraction of duplicated genes rather than by habitat variability when PICs are considered (Table 3). This result implies that the observed positive correlation between $Q_{ud}$ and habitat variability was an artifact caused by the primary association between the fraction of duplicated genes and habitat variability.

However, the effect of the fraction of duplicated genes on the increase in $Q_{ud}$ may not be conserved among living organisms, given that no positive relationship between $Q_{ud}$ and the fraction of duplicated genes was found for mammals when considering PICs ($r = 0.41$, $p = 0.18$).

## 4. Discussion

The results of this study did not confirm an association between metabolic network modularity and habitat variability, thereby rejecting the hypothesis derived from previous analytic results (Parter et al., 2007) and theoretical models (Hintze & Adami, 2008; Kashtan & Alon, 2005; Lipson et al., 2002) that habitat variability should promote network modularity. Although a positive correlation between the undirected version of network modularity and habitat variability was observed in the enzymatic networks of flies, this may be an artifact caused by the underlying known relationship between gene duplication and habitat variability in flies (Makino & Kawata, 2012). These results are consistent with the conclusions obtained from recent theoretical (Clune et al., 2013; Friedlander et al., 2013; Solé & Valverde, 2008; Takemoto, 2012b) and empirical (Takemoto & Borjigin, 2011; Takemoto, 2013) studies that cast doubt on the effect of habitat variability on network modularity. Furthermore, most of these previous studies (Takemoto & Borjigin, 2011; Takemoto, 2013) were limited to data from lower organisms such as archaea and bacteria. Therefore, the present results show that the limited impact of habitat variability on metabolic network modularity is also applicable to higher organisms such as flies and mammals. The fact that habitat variability does not generally promote metabolic network modularity may be common to all living organisms.

Moreover, these results imply that the factors that determine a species' habitat range are not consistent between a higher level of organization (i.e., biochemical network structure) and the gene level (i.e., genetic variation (Barrett & Schluter, 2008; Kellermann et al., 2009) and gene duplication (Makino & Kawata, 2012; Tamate et al., 2014)). This finding may be explained by a study conducted by Blank et al. (Blank, Kuepfer, & Sauer, 2005), who showed that network redundancy through duplicated genes was the major mechanism contributing to metabolic network robustness in yeast, with a minor contribution from alternative molecular pathways. In compound networks, network redundancy corresponds to multiple edges between a specific substrate–product pair (i.e., the number of enzymes catalyzing a specific metabolic reaction); thus, the network structure does not change in the face of gene duplication, although metabolic network robustness increases because of the high number of redundant enzymes (i.e., isozymes). On the other hand, alternative pathways resulting from gene duplication would increase network modularity (Takemoto, 2012b) because reactions are dense among a metabolite group; however, such pathways have a negligible contribution to the increase of compound network modularity because they have a minor contribution. In enzymatic networks, on the other hand, the redundancy through duplicated genes may increase metabolic network modularity. According to the definition (see Section 2.3), enzyme networks have dense subgraphs (modules) that consist of enzymes (nodes) catalyzing similar metabolic reactions, because several enzymes can share a similar substrate and/or product. Since such similar enzymes often result from gene duplication, more duplicated genes may increase the enzymatic network modularity ($Q_{ud}$). In fact, Díaz-Mejía et al. (Díaz-Mejía et al., 2007) reported a high retention of duplicates within functional modules in enzymatic networks. This fact also explains the observed positive relationship between enzymatic network modularity and the fraction of duplicated genes in flies (Figure 3). However, as described above, the effect of gene duplication on enzymatic network modularity may be not be a general phenomenon, because no positive association was observed between enzymatic network modularity and the fraction of duplicated genes. Therefore, it is likely that several factors determine metabolic network modularity.

Horizontal gene transfer is also an important evolutionary event occurring at the gene level that can be related to environmental adaptation (Kreimer, Borenstein, Gophna, & Ruppin, 2008; Pál, Papp, & Lercher, 2005); in particular, the resulting increased genetic diversity could increase the types of habitats available, and thus a species' habitat variability. However, such events are not expected to influence the metabolic network modularity of higher organisms because the major driving force of metabolic network evolution in eukaryotes, including higher organisms, is gene duplication (Pál et al., 2005; Papp et al., 2004) rather than horizontal gene transfer. Moreover, I previously reported the limited effect of horizontal gene transfer on compound and enzymatic network modularity (Takemoto, 2013) in bacteria. Note that Kreimer et al. (Kreimer et al., 2008) found a positive correlation between the degree of horizontal gene transfer and enzymatic network modularity in bacteria; however, we proposed that the observed correlation was probably due to a lack of available data on metabolic reactions (Takemoto, 2013).

Alternative theories are required to explain the origin and evolution of network modularity. The current theory (Kashtan & Alon, 2005) that habitat variability promotes network modularity was originally developed based on a genetic programming approach assuming an edge-switching mechanism; however, this basic assumption may be not satisfied. Although the rate of edge rewiring may vary slightly at each estimated divergence time, the ratio of the edge-rewiring rate in metabolic networks (Shou et al., 2011) to that in gene regulatory networks ranges from $3.2 \times 10^{-4}$ to $8.4 \times 10^{-3}$. Thus, it may be difficult to completely explain the origin and evolution of network modularity using this theory. The following alternative theories can be considered. First, the connection-cost theory (Clune et al., 2013) predicts that network modularity evolves under selection pressure to maximize network performance and minimize connection costs. Second, the multiplicative-mutation theory (Friedlander et al., 2013) predicts that multiplicative mutations lead to network modularity. Both theories may be biologically plausible because of the optimality mechanism of evolution (de Vos, Poelwijk, & Tans, 2013) and common observations of multiplicative mutations (Friedlander et al., 2013). However, more careful examinations (involving tests with empirical data) may be required. In particular, appropriate definitions of performance and connection costs will be needed to accurately test the connection-cost theory. In this context, flux balance analysis (Bordbar, Monk, King, & Palsson, 2014) and databases of the thermodynamic properties of biochemical compounds and reactions (e.g., eQuilibrator (Flamholz, Noor, Bar-Even, & Milo, 2012)) may be useful for quantifying the performance and connection costs in metabolic networks. On the other hand, it may be difficult to test the multiplicative-mutation theory because of the inherent product-rule nature of biological mutations, as the authors originally mentioned when this theory was proposed (Friedlander et al., 2013). The neutral theory for metabolic networks (Takemoto, 2012b) can quantitatively demonstrate that metabolic network modularity can arise from simple evolutionary processes, using empirical data. Therefore, the neutral theory, which predicts that network modularity can be acquired neutrally, is the most plausible hypothesis at present. The fact that metabolic networks appear to be modular in nature but not significantly so (Holme, Huss, & Lee, 2011) also supports this hypothesis.

The present results do not entirely discount the effect of habitat variability on metabolic network modularity. Rather, they highlight the need for more detailed examination of the relationships between habitat variability and metabolic networks (biological systems, in general). In particular, higher-level metabolic network analyses are needed. Although metabolic network analysis based on network science is powerful, the simplification required to visualize and represent such a network (i.e., a graph) results in several gaps in biological information (Takemoto, 2012a). In this context, the following methods may be useful: flux balance analysis (Bordbar et al., 2014) and Scope (Handorf, Ebenhöh, & Heinrich, 2005), a computational framework used to characterize the biosynthetic capability of a network when it is provided with certain external resources. In particular, the Scope algorithm has identified a functional group of metabolic networks that determines biologically important physiological parameters such as metabolic rate

(especially respiratory rate) and lifespan (Takemoto & Kawakami, 2015). These methods may help to understand the relationship between metabolic network structure and metabolic/physiological functions. In addition, the expression and activity of metabolic enzymes could also be considered using microarray and mass spectrometry.

The present analysis has some limitations that should be acknowledged. For example, the definition of modularity and modules is controvertible. Therefore, any of the conclusions reached are limited to the context of network modularity. For metabolic networks, however, biologically functional modules (i.e., functional categories of genes/proteins, curated manually) may not be accurately defined through module detection methods based on network topology (i.e., in the context of network modularity). In particular, the definition of modularity used in this study and many previous studies might not be topologically intuitive due to the locality and limited resolution (Fortunato & Barthélemy, 2007). Methods based on linked communities (Ahn, Bagrow, & Lehmann, 2010) may be more useful to avoid these limitations, because they show better accuracy in the prediction of biologically functional modules (or related categories such as pathways). In addition, it is important to consider biological information such as reaction mechanisms, the direction of reaction (i.e., reversible vs. irreversible), the chemical structure of metabolites, and gene clusters. In this context, methods for finding biologically meaningful modules of biological networks based on gene clusters and chemical transformation patterns (Kanehisa, 2013; Muto et al., 2013) may be useful.

Comparative phylogenetic analysis (i.e., PICs) also has limitations. In particular, this analysis assumed a Brownian motion-like evolution of biological traits on a phylogenetic tree with accurate branch lengths; thus, the phylogenetic analysis may result in misleading conclusions. For instance, Griffith et al. (Griffith, Moodie, & Civetta, 2003) pointed out that loss of statistical power can occur when a dataset is reduced in size owing to phylogenetic corrections. Because the present dataset contained only a few samples, it falls into the condition described by Griffith et al. However, similar conclusions were obtained when considering the phylogenetic analysis and simple correlation analysis; thus, it is expected that this limitation did not influence the conclusions.

The results of the present study also depend on the quality of the genome annotation and metabolic reaction databases. For example, opposite conclusions have been observed when using an earlier version and the latest version of metabolic networks in the same analysis (Takemoto, 2013, 2014). Furthermore, it is possible that these results were influenced by the percentage of proteins with no known function in the study organisms. For metabolic networks, the difference in the fraction of functionally unknown proteins between species categories was confirmed in a previous study of our research group (Takemoto & Yoshitake, 2013). Thus, the quality of the annotation and databases are not considered to affect the present conclusions; however, more careful examinations are required. For example, the metabolic networks themselves are not fully understood. In particular, the existence of enzyme promiscuity (Khersonsky & Tawfik, 2010), which implies that enzymes can catalyze multiple reactions, act on more than one substrate, or exert a range of suppressions (Patrick, Quandt, Swartzlander, & Matsumura, 2007), suggests the possibility of many hidden metabolic reactions, which may in turn be related to metabolic robustness against changing environments (Nam et al., 2012). Consideration of these hidden metabolic reactions will be important for designing metabolic pathways and for developing a deeper understanding of metabolic evolution.

It will also be necessary to test the relationship between habitat variability and biological systems (e.g., metabolic networks) using additional organisms. For example, it is not possible to determine whether the present conclusions are general or due to specific effects in a few individuals. Therefore, the continuous sequencing of genomes from a wide range of organisms (e.g., mammals, birds, fish, and insects) is clearly important for this field to progress. Further development of high-throughput sequencing techniques will enable the rapid collection of such data. For example, metagenomics techniques are now available to complete the sequencing of an organism's genome.

Despite the limitations of the present analysis mentioned above, these findings nonetheless encourage a reconsideration of the widely accepted hypothesis that habitat variability promotes network modularity. Furthermore, these results enhance general understanding of adaptive and evolutionary mechanisms (i.e., mechanisms for habitat variability) in metabolic networks.

## Acknowledgements

This study was supported by a Grant-in-Aid for Young Scientists (A) from the Japan Society for the Promotion of Science (no. 25700030). KT thanks J.-B. Mouret for providing an executable file for calculating $Q_d$.

## Tables

**Table 1: Correlations between habitat variability (Brillouin index) and modularity of compound networks.**
$Q_d$ and $Q_{ud}$ indicate the modularity scores for directed networks and undirected networks, respectively. Correlation analyses were performed with and without consideration of phylogenetically independent contrasts (PICs). Pearson's correlation coefficient *r* is shown. Parenthetic values indicate the associated *p*-values.

| Species category | Flies | | Mammals | |
|---|---|---|---|---|
| Modularity | PICs Considered | PICs Not Considered | PICs Considered | PICs Not Considered |
| $Q_d$ | –0.56 (0.094) | –0.32 (0.33) | –0.37 (0.23) | 0.24 (0.42) |
| $Q_{ud}$ | –0.61 (0.063) | –0.42 (0.20) | –0.45 (0.14) | –0.27 (0.36) |

**Table 2: Correlations between habitat variability (Brillouin index) and modularity of enzymatic networks.**
See Table 1 for description of the method of calculation and table elements.

| Species category | Flies | | Mammals | |
|---|---|---|---|---|
| Modularity | PICs Considered | PICs Not Considered | PICs Considered | PICs Not Considered |
| $Q_d$ | –0.70 (0.023) | 0.13 (0.71) | 0.17 (0.61) | 0.59 (0.035) |
| $Q_{ud}$ | 0.91 (0.00031) | 0.78 (0.0042) | –0.059 (0.86) | –0.31 (0.30) |

**Table 3: Influences of habitat variability and fraction of duplicated genes on the modularity ($Q_{ud}$) of enzymatic networks in flies.**
SE indicates the standard error.

| Variable | PICs Considered | | | PICs Not considered | | |
|---|---|---|---|---|---|---|
| | Estimate | SE | *P*-value | Estimate | SE | *P*-value |
| Habitat variability | –0.0023 | 0.0024 | 0.38 | 0.0060 | 0.0043 | 0.20 |
| Fraction of duplicated genes | 0.83 | 0.13 | 0.00041 | 0.30 | 0.21 | 0.20 |

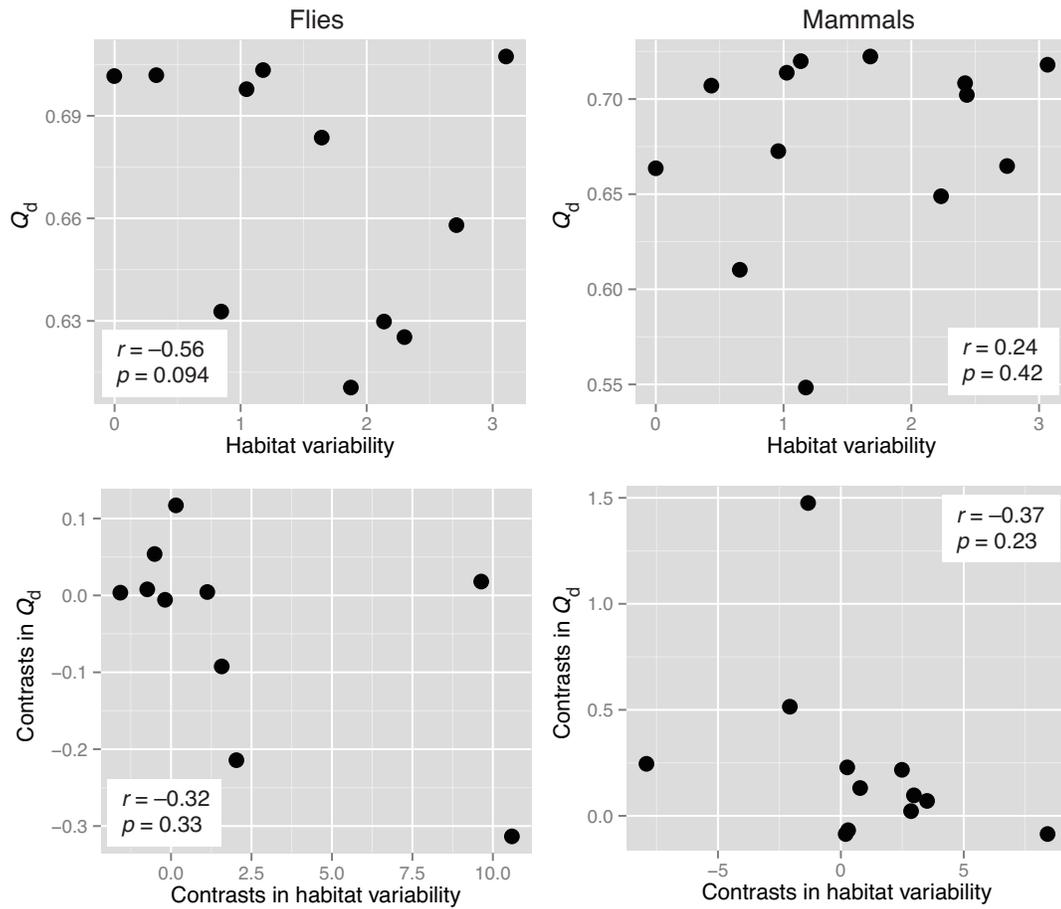

**Figure 1: Relationships between the modularity score for directed networks ($Q_d$) and habitat variability.**
The horizontal axes represent habitat variability (Brillouin index) (top panels) and phylogenetically independent contrasts (PICs) in the index (lower panels). The vertical axes indicate $Q_d$ (top panels) and PICs in $Q_d$ (bottom panels) of compound networks in flies (left panels) and mammals (right panels). Pearson's correlation coefficients ($r$) and associated *p*-values are shown.

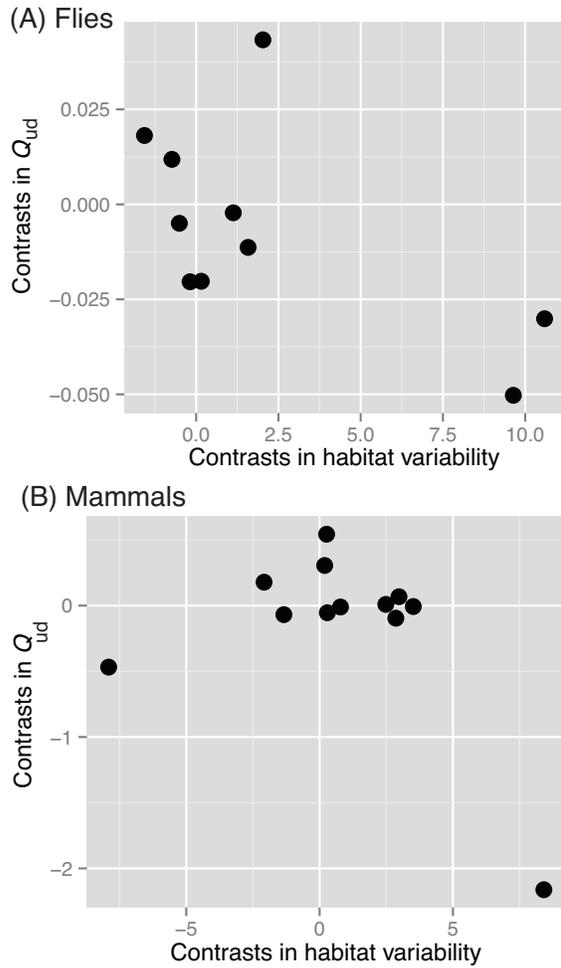

**Figure 2: Relationship between modularity score for undirected networks ($Q_{ud}$) and habitat variability.**
The horizontal and vertical axes indicate phylogenetically independent contrasts (PICs) in habitat variability (Brillouin index) and PICs in the $Q_{ud}$ of compound networks, respectively, in flies (A) and mammals (B). A weak negative correlation and no correlation were observed for flies ($r = -0.61$, $p = 0.063$) and mammals ($r = -0.45$, $p = 0.14$), respectively.

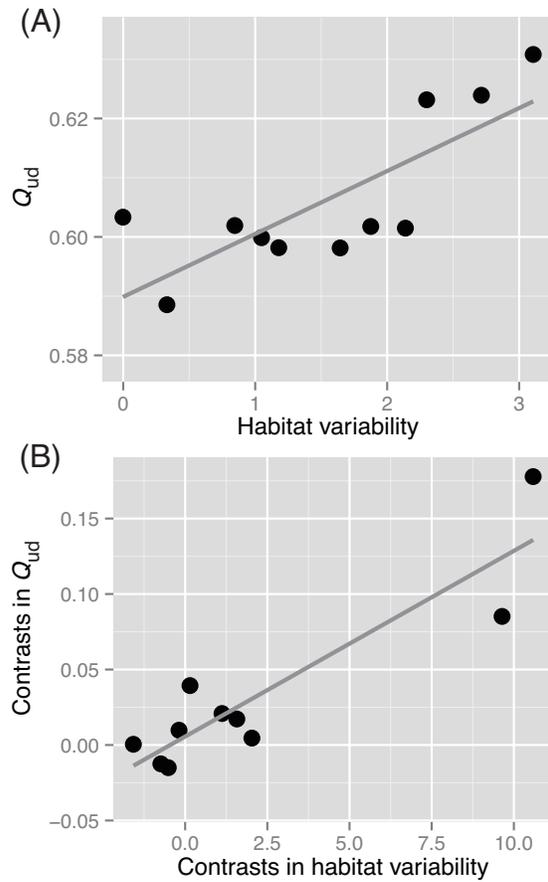

**Figure 3: Correlation between modularity score ($Q_{ud}$) and habitat variability in the enzymatic networks of flies.**
A positive correlation was observed both when phylogenetically independent contrasts (PICs) were not considered (A; $r = 0.78$, $p = 0.042$) and when the PICs were considered (B; $r = 0.91$, $p = 3.1 \times 10^{-3}$).

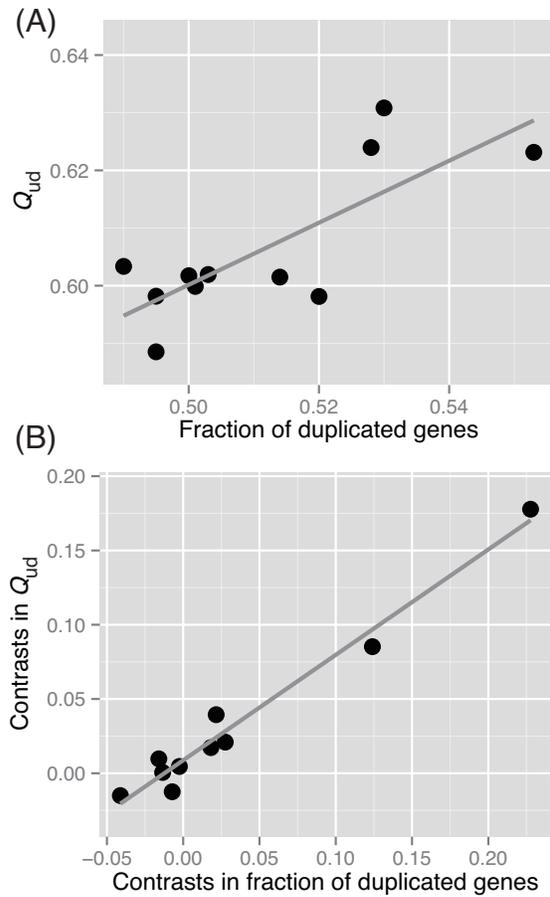

**Figure 4: Correlation between modularity score ($Q_{ud}$) and the fraction of duplicated genes in the enzymatic networks of flies.**
A positive correlation was observed both when phylogenetically independent contrasts (PICs) were not considered (A; $r = 0.78$, $p = 0.0043$) and when the PICs were considered (B; $r = 0.98$, $p = 2.4 \times 10^{-7}$).